\documentclass[twoside,fleqn]{article}
\usepackage{espcrc2,epsfig,amsmath,array}

\DeclareMathOperator{\diag}{diag}
\makeatletter
\def\thickhline{%
  \noalign{\ifnum0=`}\fi\hrule \@height 1pt \futurelet
   \reserved@a\@xhline}
\makeatother

\title{Beyond the Thouless energy}

\author{M.E.~Berbenni-Bitsch\address{Fachbereich Physik--Theoretische
    Physik, Universit\"at Kaiserslautern, D-67663 Kaiserslautern,
    Germany}, 
  M.~G\"ockeler\address{Institut f\"ur Theoretische Physik,
    Universit\"at Regensburg, D-93040 Regensburg, Germany},
  H.~Hehl$^{\rm b}$,
  S.~Meyer$^{\rm a}$,
  P.E.L.~Rakow$^{\rm b}$,
  A.~Sch\"afer$^{\rm b}$, and
  T.~Wettig\address{Institut f\"ur Theoretische Physik, Technische
    Universit\"at M\"unchen, D-85747 Garching,
    Germany}}

\begin{document}

\begin{abstract}
  The distribution and the correlations of the small eigenvalues of
  the Dirac operator are described by random matrix theory (RMT) up to
  the Thouless energy $E_c\propto 1/\sqrt{V}$, where $V$ is the
  physical volume.  For somewhat larger energies, the same quantities
  can be described by chiral perturbation theory (chPT).  For most
  quantities there is an intermediate energy regime, roughly
  $1/V<E<1/\sqrt{V}$, where the results of RMT and chPT agree with
  each other.  We test these predictions by constructing the connected
  and disconnected scalar susceptibilities from Dirac spectra obtained
  in quenched SU(2) and SU(3) simulations with staggered fermions for
  a variety of lattice sizes and coupling constants.  In deriving the
  predictions of chPT, it is important to take into account only those
  symmetries which are exactly realized on the lattice.
\end{abstract}

\maketitle

The theoretical understanding of the Dirac eigenvalue spectrum in a
finite volume has improved considerably in recent years.  The smallest
Dirac eigenvalues are described by universal functions which can be
computed most easily in chiral RMT \cite{Verb99,Wett99}.  The
agreement persists up to the so-called Thouless energy $E_c$ which
scales like $1/L^2$, where $V=L^4$ \cite{Jani98,Osbo98,Berb98b}.
Beyond this energy, the Dirac spectrum can be described by chPT
\cite{chPT}.  This has been discussed in the continuum theory in
Ref.~\cite{Osbo99}.  On a coarse lattice, the situation is different,
and one should take into account only the lattice symmetries.  Here,
we present an analysis appropriate for staggered fermions at
relatively strong coupling and compare our predictions to SU(2) and
SU(3) lattice gauge data.  For details of the SU(2) analysis, we refer
to Ref.~\cite{Berb99}.

We are interested in the connected and disconnected scalar
susceptibilities defined by
\begin{subequations}
  \label{susc}
  \begin{eqnarray}
    \chi^{\mathrm {conn}}_{\mathrm {lat}}(m)
    &\!\!=\!\!& -\frac1N\left\langle\sum_{k=1}^N
      \frac1{(i\lambda_k+m)^2}\right\rangle\:,\\
    \chi^{\mathrm {disc}}_{\mathrm {lat}}(m)
    &\!\!=\!\!&\frac{1}{N}\left\langle\sum_{k,l=1}^N \frac{1}
      {(i\lambda_k+m)(i\lambda_l+m)}\right\rangle \nonumber\\
    && -\frac{1}{N} \left\langle\sum_{k=1}^N\frac{1}
      {i\lambda_k+m}\right\rangle^2 \,,
  \end{eqnarray}
\end{subequations}
respectively, where the $i\lambda_k$ are the Dirac eigenvalues and $m$
is a valence quark mass.  Most of the RMT-predictions for these
quantities are given in Refs.~\cite{Berb99,Gock99}.  The corresponding
chPT-predictions can be derived from an effective partition function
$Z$ by differentiating with respect to the quark masses \cite{Berb99}.
We consider $N_v$ generations of valence quarks of mass $m_v$ and
$N_s$ generations of sea quarks of mass $m_s$ (corresponding to $4
N_v$ valence quarks and $4 N_s$ sea quarks in the continuum limit).
Our starting point is the following expression for the free energy,
\begin{eqnarray}
  \label{part}
  \ln Z(m_v,m_s,L) \propto  \:VS(m_v,m_s)\hspace*{20mm}\nonumber\\
  -\frac12\sum_QK_Q\sum_p \ln\left[\hat{p}^2+m_Q^2(m_v,m_s)\right]\:,
\end{eqnarray}
where $S(m_v,m_s)$ is the saddle-point contribution, and the double
sum represents the one-loop contribution coming from light composite
bosons.  The sum runs over the allowed momenta $p$ ($p_\mu = 2 \pi
n_\mu /L$ with integer $n_\mu$) and over particle type $Q$ (with
multiplicity $K_Q$ and mass $m_Q$).  We use ${\hat p}^2
\equiv2\sum_\mu(1-\cos p_\mu)$.

The main task is to determine the $K_Q$ and $m_Q$ for our particular
problem.  Consider first gauge group SU(3).  The symmetry in the
chiral limit is
$\text{SU}(N_v+N_s)\times\text{U}(1)\times\text{SU}(N_v+N_s)
\times\text{U}(1)$ which is spontaneously broken to
$\text{SU}(N_v+N_s)\times\text{U}(1)$.  Since for staggered fermions
in strong coupling the U(1) symmetry is anomaly-free, we expect
$(N_v+N_s)^2$ Goldstone bosons.  The bosons made of different quark
flavors $\bar q_i q_j$ will have a mass given by $m^2=A(m_i+m_j)/2$.
(According to the Gell-Mann--Oakes--Renner relation,
$A=2\Sigma/f_\pi^2$, where $\Sigma=|\langle\bar\psi\psi\rangle|$.)  We
thus have $N_v^2-N_v$ mesons of mass $Am_v$, $N_s^2-N_s$ mesons of
mass $Am_s$, and $2N_vN_s$ mesons of mass $A(m_v+m_s)/2$.

\begin{figure}[-t]
  \centerline{\epsfig{figure=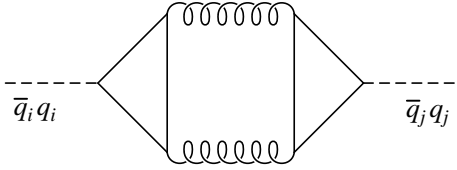,width=60mm}}
  \vspace*{-6mm}
  \caption{Annihilation diagram for the ``flavor-diagonal'' mesons.}
  \label{fig1}
  \vspace*{-4mm}
\end{figure}

For ``flavor-diagonal'' mesons we must also consider the annihilation
process in Fig.~\ref{fig1}.
Because of the anomaly-free U(1) symmetry, the amplitude for $\bar
q_iq_i\to\bar q_jq_j$ vanishes for $m_i=0$ or $m_j=0$.  Therefore, we
make the following ansatz for the mass-squared matrix of the states
$(\bar v_1v_1,\dots,\bar v_{N_v}v_{N_v}, \bar s_1s_1,\dots,\bar
s_{N_s}s_{N_s})^T$,
\begin{eqnarray*}
  M^2=A\diag(m_v,\dots,m_v,m_s,\dots,m_s)\hspace*{16mm}\\  
  +z\begin{pmatrix}
    m_v^2&\!\!\!\cdots\!\!\!&m_v^2&m_vm_s&\!\!\!\cdots\!\!\!&m_vm_s\\
    \vdots&&\vdots&\vdots&&\vdots\\
    m_v^2&\!\!\!\cdots\!\!\!&m_v^2&m_vm_s&\!\!\!\cdots\!\!\!&m_vm_s\\
    m_sm_v&\!\!\!\cdots\!\!\!&m_sm_v&m_s^2&\!\!\!\cdots\!\!\!&m_s^2\\
    \vdots&&\vdots&\vdots&&\vdots\\
    m_sm_v&\!\!\!\cdots\!\!\!&m_sm_v&m_s^2&\!\!\!\cdots\!\!\!&m_s^2
  \end{pmatrix}
\end{eqnarray*}
with an additional parameter $z$.  The eigenvalues of $M^2$ are $Am_v$
with multiplicity $N_v-1$, $Am_s$ with multiplicity $N_s-1$, and
$\lambda_\pm$ with multiplicity one (the expression for $\lambda_\pm$
is given in \cite{Berb99}).  This completes the determination of the
light boson spectrum of the gauge group SU(3) in Table~\ref{table1}.

\begin{table}[-t]
  \begin{center}
    \renewcommand{\arraystretch}{1.5}
    \begin{tabular}{!{\vrule width 1pt}c!{\vrule width 1pt}c|
        c!{\vrule width 1pt}}\thickhline & \multicolumn{2}
      {c!{\vrule width 1pt}}{multiplicity} \\\cline{2-3} 
      \raisebox{1.5ex}[0pt][0pt]{$m^2$} & SU(2) & 
      \hspace*{8pt} SU(3) \hspace*{8pt}\\\thickhline
      $Am_v$ & $2N_v^2+N_v-1$ & $N_v^2-1$ \\\hline
      $Am_s$ & $2N_s^2+N_s-1$ & $N_s^2-1$ \\\hline
      $A(m_v+m_s)/2$ & $4N_vN_s$ & $2N_vN_s$ \\\hline
      $\lambda_-$ & 1 & 1 \\\hline
      $\lambda_+$ & 1 & 1 \\\thickhline
    \end{tabular}  
  \end{center}
  \caption{The light particle spectrum for gauge groups SU(2) and SU(3).}
  \label{table1}
  \vspace*{-5mm}
\end{table}

For the gauge group SU(2), the symmetry in the chiral limit is
$\text{U}(2N_v+2N_s)$, spontaneously broken to
$\text{O}(2N_v+2N_s)$~\cite{Klub83}.  We thus have
$(N_v+N_s)(2N_v+2N_s+1)$ Goldstone particles.  Some of the baryon
($q_i q_j$ and $\bar q_i\bar q_j$) states have the same mass as the
mesons, $m^2=A(m_i+m_j)/2$, giving rise to the light particle spectrum
in Table~\ref{table1}.

Table~\ref{table1} determines the one-loop contribution to the free
energy in Eq.~(\ref{part}).  The saddle-point contribution is
parameterized by a smooth function of $m_v$ and $m_s$, independent of
the lattice size.  Taking appropriate derivatives of $\ln Z$ with
respect to the quark masses \cite{Berb99}, we obtain the
chPT-predictions for the susceptibilities of Eq.~(\ref{susc}).  In the
final results, we take the limits $m_v=m_s=m$, $N_v\to0$, and
$N_s\to0$.  The fit parameters are $A$, $z$, and the smooth
background.  Since $\Sigma$ can be determined independently by a fit
to RMT, our results for the parameter $A=2\Sigma/f_\pi^2$ also give us
an estimate of $f_\pi$ \cite{Berb99}.

Taking the infinite-volume limit of the chPT-expressions, we obtain
several terms containing logarithms in the quark mass \cite{Berb99}.
Note, however, that the leading term $\propto \ln m$ in the chiral
condensate, which is expected in the quenched approximation
\cite{chPT}, is absent in our case because of the anomaly-free U(1)
symmetry.

\begin{figure}[-t]
  \centerline{\epsfig{figure=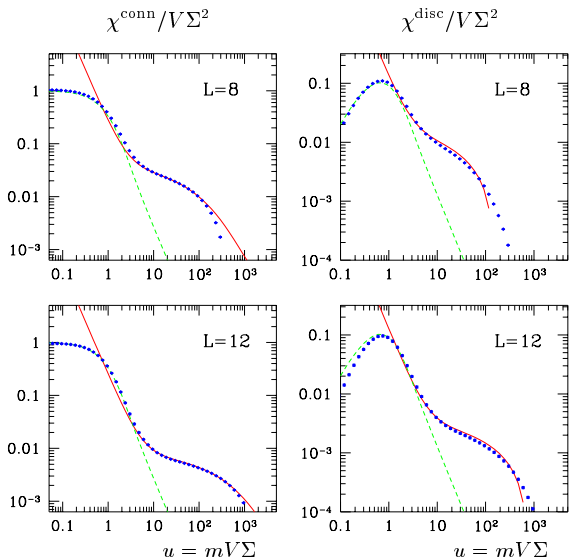,width=75mm}}
  \vspace*{-8mm}
  \caption{Connected and disconnected scalar susceptibilities versus the
    rescaled valence quark mass for staggered fermions using gauge
    group SU(2) at $\beta=4/g^2=2.2$ and $V=8^4$ and $12^4$.}
  \label{fig2}
  \vspace*{-6mm}
\end{figure}

\begin{figure}[-t]
  \centerline{\epsfig{figure=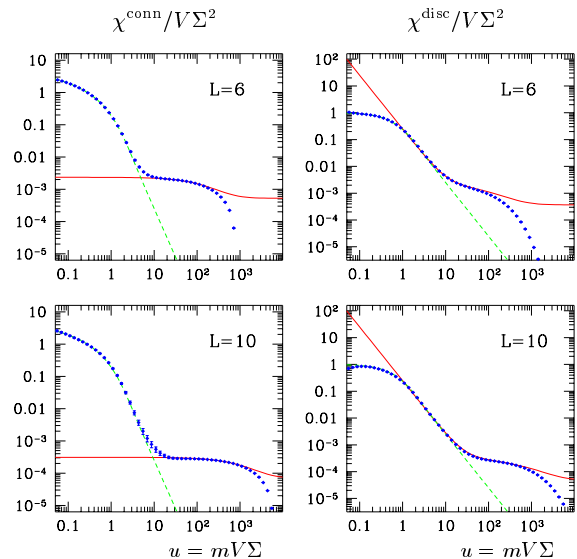,width=75mm}}
  \vspace*{-8mm}
  \caption{Same as Fig.~\protect\ref{fig2} but for gauge group SU(3)
    at $\beta=6/g^2=5.4$ and $V=6^4$ and $10^4$.}
  \label{fig3}
  \vspace*{-3mm}
\end{figure}

Our results for gauge group SU(2) and SU(3) are displayed in
Figs.~\ref{fig2} and \ref{fig3}.  The diamonds represent the lattice
data plotted vs.~the rescaled valence quark mass $u=mV\Sigma$, the
solid lines the (finite-volume) chPT predictions, and the dashed lines
the RMT predictions (for topological charge $\nu=0$), respectively.
As expected, for $u<f_\pi^2L^2$ the data are described by RMT.  For
$u>1$, they are very well described by our chPT expressions. (chPT
breaks down for $u<1$ since the $p=0$ modes must be treated
non-perturbatively in this region.  The deviations between lattice
data and chPT for very large $u$ are 
due to the finite
lattice.)  The domain of common applicability of RMT and chPT,
$1<u<f_\pi^2L^2$, grows with the lattice size.

In the case of the connected susceptibility in SU(3) (see
Fig.~\ref{fig3}) we do not see an overlap region of RMT and chPT.  The
reason is that for this particular quantity (and also for the chiral
condensate) the would-be leading terms both in RMT (for large $m$) and
in chPT (for small $m$) are absent.  This is a rather special case
caused by the anomaly-free U(1) symmetry and by the fact that $N_s=0$.
As a consequence, the Thouless energy for this quantity scales like
$1/L^{8/3}$ instead of like $1/L^2$ so that RMT breaks down for
$u\propto L^{4/3}$.

In conclusion, we now have a good theoretical understanding of the
finite-volume Dirac spectrum also beyond the Thouless energy.  Our
analysis was tailored to the case of staggered fermions at strong
coupling where the anomaly-free U(1) symmetry causes the light particle
spectrum to be different from that of the continuum theory.

We thank M.~Golterman and J.J.M.~Ver\-baar\-schot for helpful comments.
This work was supported in part by DFG.

\end{document}